\documentstyle[12pt]{article}

\normalbaselines
\catcode `\@=11
\@addtoreset{equation}{section}

\def\section{\@startsection {section}{1}{\z@}{-3.5ex plus -1ex minus
     -.2ex}{2.3ex plus .2ex}{\normalsize\bf}}
\def\subsection{\@startsection{subsection}{2}{\z@}{-3.25ex plus -1ex minus
 -.2ex}{1.5ex plus .2ex}{\normalsize\bf}}
\def\@cite#1#2{${}^{\mbox{\scriptsize#1\if@tempswa , #2\fi}}$}
\def\thebibliography#1{\section*{References\markboth
  {REFERENCES}{REFERENCES}}\list
  {\arabic{enumi}.}{\settowidth\labelwidth{[#1]}\leftmargin\labelwidth
  \advance\leftmargin\labelsep
  \usecounter{enumi}}
  \def\newblock{\hskip .11em plus .33em minus -.07em}
  \sloppy
  \sfcode`\.=1000\relax}
 
\catcode `\@=12

\newcommand{\pnds}{\pounds}
\newcommand{\beq}{\begin{equation}}
\newcommand{\eeq}{\end{equation}}
\newcommand{\beqa}{\begin{eqnarray}}
\newcommand{\eeqa}{\end{eqnarray}}
\newcommand{\nn}{\nonumber}

\newcommand{\half}{\frac{1}{2}}

\newcommand{\xt}{\tilde{X}}

\newcommand{\uind}[2]{^{#1_1 \, ... \, #1_{#2}} }
\newcommand{\lind}[2]{_{#1_1 \, ... \, #1_{#2}} }

\newcommand{\co}{\circ}

\newcommand{\pbr}[2]{ \{ \hspace*{-2.2pt} [ #1 , #2 ] \hspace*{-2.5pt} \} }
\newcommand{\we}{\wedge}
\newcommand{\dv}{d^V}
\newcommand{\nbrpq}[2]{\nbr{\xxi{#1}{1}}{\xxi{#2}{2}}}

\newcommand{\rbox}[2]{\raisebox{#1}{#2}}
\newcommand{\xx}[1]{\raisebox{1pt}{$\stackrel{#1}{X}$}}
\newcommand{\xxi}[2]{\raisebox{1pt}{$\stackrel{#1}{X}$$_{#2}$}}

\newcommand{\ff}[1]{\raisebox{1pt}{$\stackrel{#1}{F}$}}

\newcommand{\nbr}[2]{{\bf[}#1 , #2{\bf ]}}

\newcommand{\der}{\partial}

\newcommand{\Om}{\Omega}
\newcommand{\om}{\omega}

\newcommand{\inn}{\hspace*{2pt}\raisebox{-1pt}{\rule{6pt}{.3pt}\hspace*
{0pt}\rule{.3pt}{8pt}\hspace*{2pt}}}

\newcommand{\bm}{\boldmath}
\newcommand{\bd}{\mbox{\bm $d$}}

\newcommand{\lapl}{\bigtriangleup}
\newcommand{\psib}{\bar{\psi}}
\newcommand{\derts}{\stackrel{\leftrightarrow}{\der}}

\begin{document}

\begin{flushright}
hep-th/9511039
\end{flushright}

\vspace*{2.5cm}
\noindent
{\bf FROM THE POINCAR\'E--CARTAN FORM TO A GERSTENHABER \\
ALGEBRA OF  POISSON BRACKETS
IN FIELD   THEORY{}\footnote{
submitted Jan. 1995, revised July 1995;
to appear in
{\em Quantization, Coherent States and Complex Structures,}
%%{\rm Proc. of the XIII-th Int. Workshop
%%on Geometric Methods in Physics},
%%Bia\l owie$\dot {\rm z}$a, Poland, July 1994,
S.T. Ali, J.P. Antoine e.a. eds., Plenum Press, N.Y. 1995}
}
\vspace{1.3cm}\\
\noindent
\hspace*{1in}
%\begin{minipage}{13cm}
\begin{center}
Igor V. Kanatchikov\footnote{New address: Center of Theoretical
Physics, Polish Academy of Sciences,
al. Lotnikov 32/46,
PL-02-668 Warsaw, Poland}  \vspace{0.3cm}\\
Institut f\"ur Theoretische Physik  \\RWTH Aachen \\
D-52056 Aachen, Germany \\
%\end{minipage}
\end{center}

\vspace*{0.5cm}

\begin{abstract}
\noindent
We consider the generalization of the basic structures of classical
analytical mechanics to field theory within the framework of the
De Donder-Weyl (DW) covariant canonical theory. We start from the
Poincar\'e-Cartan form and construct the analogue of the symplectic
form -- the polysymplectic form of degree $(n+1)$, $n$ is the dimension
of  the space-time. The dynamical variables are represented by
differential forms and the polysymplectic form leads to a natural
definition of the Poisson brackets on forms. The Poisson brackets
equip the exterior algebra  of dynamical variables with the structure
of a  "higher-order" Gerstenhaber algebra. We also briefly
discuss a possible approach to
field quantization  which proceeds from the DW Hamiltonian formalism
and the Poisson brackets of forms.
\end{abstract}

\section{\hspace{-4mm}.\hspace{2mm}INTRODUCTION }

\hspace{0.8cm} In this communication I discuss the canonical structure
underlying the so-called De Donder--Weyl (DW) Hamiltonian
formulation in field theory and its possible application
to a quantization of fields. The abovementioned structure
was found in a recent paper of mine,\cite{Kan} to which I refer
both for further references and for additional details.
In particular, I am going to show that the relationships
between the Poincar\'e-Cartan form, the symplectic structure
and the Poisson
structure, which are well known in the mathematical formalism of
classical mechanics, have their natural counterparts
%analogues
also in field theory within
the framework of the DW canonical theory.
This leads to the analogue of the symplectic structure,
which I call polysymplectic, and to the analogue
of the Poisson brackets which are defined on
differential forms.

Recall that the
Euler-Lagrange field equations may be
written in the following form
(see for instance Refs. 2-4) %%\cite{Rund,Kastrup83,Binz})
\beq
\frac{\der p^i_a}{\der x^i}=-\frac{\der H}{\der y^a},
\hspace*{50pt}
\frac{\der y^a}{\der x^i}=\frac{\der H}{\der p^i_a}
\eeq
%\nopagebreak[3]
in terms of the variables
%\nopagebreak[3]
\beqa
p^i_a &:=& \frac{\der L}{\der(\der_i y^a)} , \\
H&:=&p^i_a\der_iy^a -L
\eeqa
%\pagebreak[3]\\
which are to be refered to as the DW momenta and
the DW Hamiltonian function respectively.
Here $L=L(y^a,\der_i y^a, x^i)$ is the Lagrangian
density, $x^i, i=1,...,n$ are   space-time
coordinates and $y^a, a=1,...,m$ are field variables.
Eqs. (1) are reminiscent of  Hamilton's canonical
equations
in mechanics and, therefore, may be thought of as a
specific covariant Hamiltonian formulation of field
equations. We call Eqs. (1) the DW
%canonical or
Hamiltonian field equations and the formulation of field
theory in terms of the variables $p_a^i$ and $H$ above
the DW Hamiltonian formulation. The formulation above
%
%the following line differs from that published in Plenum!
%
originates from the works of De Donder (1935) and Weyl (1935)
on the variational calculus of multiple integrals.

The mathematical structures underlying this formulation of
field theory were considered  earlier by several authors
in the context of
the so-called multisymplectic
for\-ma\-lism\cite{Kij ea}
which was recently studied in detail
in Refs. 6-8. %%\cite{Gotay, Gimmsy, Crampin}.
However, the possible analogues of the symplectic
structure and the Poisson brackets, which are
known to be so fruitful in the canonical formulation
of classical mechanics, still are not properly understood
within the DW canonical theory.

Our interest to this
subject is motivated  by the explicit covariance
of the formulation above, in the sense that the space and
time variables are not discriminated as usual,
and its finite dimensionality, in the sense that the
formulation refers to the finite dimensional analogue
of the phase space namely,  the space of variables
$(y^a, p_a^i)$, as well as by the attempts to understand
if or how it is possible to construct a
formulation of quantum field theory which would be  based
on the DW Hamiltonian formulation.
Clearly, the answer
to  the latter question requires the analogue of the
Poisson brackets  and the bracket representation of
the equations of motion corresponding to the DW
formulation.

The canonical formalism in classical mechanics is
related to
the variational principle of least action and it may be
derived from the fundamental object of the calculus of
variations -- the Poincar\'e-Cartan (P-C) form
(see e.g. Ref. 9).
%%\cite{Abr+Marsden}).
The corresponding construction leads to  structures which
are known to be important for quantization.
Conventional generalization to field theory implies
%specification of the time dimension
setting off the time dimension from other space-time
dimensions
and leads to the infinite dimensional
functional version of the abovementioned construction.
Here we are interested in the field theoretical generalization
of these structures within the space-time symmetric
DW formulation.

\section{\hspace{-4mm}.\hspace{2mm}POINCAR\'E--CARTAN FORM, CLASSICAL
EXTREMALS AND THE POLYSYMPLECTIC FORM }
 %P-C

\hspace{0.8cm} In field theory, which is related to the
variational problems with several independent variables,
the analogue of the P-C form written in terms of the DW
Hamiltonian variables (1.2), (1.3)
reads\cite{Gotay}$^-$\cite{Crampin}
%(we confine ourselves to the first order theories)
\beq
\Theta=p^i_a dy^a\we \om_i - H\omega ,
\eeq
where $\om:=dx^1\wedge...\wedge dx^n $
and $\om_i:=\der_i \inn \om$.
The equations of motion in the DW Hamiltonian form,
Eqs. (1.1), may be shown to follow from the
statement that the
%\nopagebreak[3]
classical extremals are the integral
hypersurfaces of the multivector field of degree $n$,
$\xx{n}$,
%\nopagebreak[3]
\beq
\mbox{$\xx{n}:=\frac{1}{n!} X$$^{M_1...M_n}(z)\,\der
\lind{M}{n} , $}
\eeq
where
%\nopagebreak[3]
$ \der \lind{M}{n} := \partial_{M_1}
\wedge...\wedge\partial_{M_n} , $
which annihilates the canonical $(n+1)$-form
%\nopagebreak[3]
\beq
\Om_{DW}:=d\Theta,
\eeq
%\pagebreak[3]
that is
\beq
\xx{n}\inn\ \Om_{DW}=0 .
\eeq
The integral hypersurfaces of $\xx{n}$ are defined
as the solutions of the equations
\begin{equation}
\xx{n}{}^{M_1...M_n}(z)=
{\cal N}
 \frac{\partial(z^{M_1},...,z^{M_n})}
{\partial(x^1,...,x^n)}
\end{equation}
where a multiplier ${\cal N}$ depends on the chosen
parametrization of a hypersurface and
$z^M := (x^i, y^a, p^i_a)$.
The component calculations show that Eq. (2.4)
specifies only a part of the components of $\xx{n}$
and that the DW canonical equations (1.1) actually follow from
the ``vertical'' components $X^v{}\uind{i}{n-1}$.
We call vertical the field and the DW momenta variables
$z^v:=(y^a,p^i_a)$ and horizontal the space-time
(independent) variables $x^i$.

%\medskip

Introducing the notions of the vertical multivector field
of degree $p$:
\beq
\xx{p}{}^V:=  \frac{1}{(p-1)!} X^v{}\uind{i}{p-1}
\der_v{}\lind{i}{p-1}  ,
\eeq
the vertical exterior differential, $\dv$,
$ \dv ... := dz^v\we\der_v \, ... \, ,  $
and the form
\beq
\Omega := -dy^a\we dp^i_a \we \om_i,
\eeq
one may check that (2.4) is equivalent to
\beq
\xx{n}{}^V \inn\ \Omega  = (-)^n d^V H,
\eeq
if the parametrization in (2.5)
is chosen such that
\[\frac{1}{n!} \xx{n}{}\uind{i}{n}\der \lind{i}{n}
\inn\ \om = 1 . \]
The form $\Omega$ in (2.7) is to be referred to as
{\em polysymplectic}.

The appearance of the DW field equations in the form of (2.8)
suggests (cf. with mechanics!) that the polysymplectic form is
a field theoretical analogue of the symplectic form, so
that its properties should be taken seriously as a starting
point for the canonical formalism.

%\medskip

As a generalization of (2.8), it is easy to see that
the polysymplectic form maps in general
the horizontal $q$-forms, $\ff{q}$,
\beq
\ff{q}:=\frac{1}{q!}F_{i_1 ... i_q}(z) dx\uind{i}{q},
\eeq
where
\[dx\uind{i}{q} := dx^{i_1}\wedge ...\wedge dx^{i_q}  ,\]
to the vertical multivectors of degree $(n-q)$:
\beq
\xx{n-q}\inn\ \Om = d^V \ff{q}
\eeq
for all $q= 0, ..., n-1$. Evidently, the horizontal forms
play a role of dynamical variables within the present
formalism. Henceforth we omit the superscripts ${}^V$
labelling the vertical multivectors.

The hierarchy of maps (2.10) may be viewed as a local consequence
of the hierarchy of ``graded canonical symmetries''
\beq
\mbox{$\pnds$\rbox{1pt}{$_{\stackrel{p}{X}}$}
$\Omega = 0 ,$}
\eeq
$p=1,...,n$, which are formulated in terms of the generalized
Lie derivatives with respect to the vertical multivector
fields. By definition,\cite{tulcz}
for any form $\mu$
\beq
\mbox{$\pnds$\rbox{1pt}{$_{\stackrel{p}{X}}$} $\mu :=$
\rbox{1pt}{$\stackrel{p}{X}$}\inn\
$d^V\mu -(-1)^p \, d^V$(\rbox{1pt}{$\stackrel{p}{X}$}
\inn\ $\mu)$}.
\eeq
Now, by analogy with the  terminology known from mechanics,
I call the
vertical multivector fields fulfilling (2.11)
{\em locally Hamiltonian}
and those fulfilling (2.10) (globally) {\em Hamiltonian}.
Correspondingly,
the forms to which the Hamiltonian multivector fields
can be associated
through the map (2.10) are referred to as the
{\em Hamiltonian forms}.

%\medskip

The notion of a Hamiltonian form implies certain
restriction on the dependence of its components on
the DW momenta. For example, the components of the vector
field $X_F:=X^a\der_a + X^i_a \der^a_i$
associated through the map $X_F\inn \Omega = \dv F$
with the $(n-1)$--form
$F:=F^i \om_i$ are given by
\beq
X^i_a=\der_aF^i ,  \quad
 -X^a\delta^i_j=\der^a_j F^i  .
\eeq
The latter relation restricts the  admissible
$(n-1)$--forms to those which have a simple dependence on the
DW momenta namely,
$F^i(y,p,x)= f^a(y,x)\, p^i_a + g^i(y,x)$.

%\medskip

Note also that the Hamiltonian multivector field
associated with a form through the map (2.10)
is actually defined up to an addition
of  {\em primitive} fields which annihilate the
polysymplectic form
\beq
\xx{p}_0\inn\ \Om=0 .
\eeq
Therefore, the image of a Hamiltonian form under the
map (2.10) given by the polysymplectic form is rather the
equivalence class of Hamiltonian multivector fields
of corresponding degree modulo an addition of primitive
fields.

\section{\hspace{-4mm}.\hspace{2mm}THE POISSON
BRACKETS ON FORMS AND A GERSTENHABER ALGEBRA }

\hspace*{0.8cm} It is natural to define the bracket of two locally
Hamiltonian multivector fields as follows:
\beq
\mbox{\nbrpq{p}{q}} \inn\ \Omega :=
\mbox{$\pnds$\rbox{1pt}{$_{\stackrel{p}{X}{}_1}$}}
%\lieni{p}{1}
(\xx{q}{}_2 \inn\ \Omega) .%}
\eeq
{}From the definition it follows that
\beqa
\mbox{$deg($\nbrpq{p}{q})} &=& p+q-1,    \\
\mbox{\nbrpq{p}{q}} &=& -(-1)^{(p-1)(q-1)}
\nbr{\xxi{q}{2}}{\xxi{p}{1}} ,   \\
\mbox{$(-1)^{g_1 g_3}$\nbr{\xx{p}}{\nbr{\xx{q}}{\xx{r}}}}
&+& %\mbox{\hspace*{15em}}
\mbox{$(-1)^{g_1 g_2}$\nbr{\xx{q}}{\nbr{\xx{r}}{\xx{p}}}} \nn \\
&+&\mbox{$(-1)^{g_2 g_3}$ \nbr{\xx{r}}{\nbr{\xx{p}}{\xx{q}}} $=0,$}
\eeqa
where $g_1=p-1, \; g_2=q-1$ and $g_3=r-1$.

These properties allow us to identify the bracket in (3.1)
with the vertical (i.e. taken w.r.t. the vertical variables)
Schouten--Nijenhuis (SN) bracket of
multivector fields and to conclude that the space of
LH fields is
a graded Lie algebra with respect to the (vertical) SN bracket.

For two Hamiltonian multivector fields one obtains
\begin{eqnarray}
\mbox{\nbrpq{p}{q}\inn\ $\Omega$} & =\, &
\mbox{\mbox{$\pnds$\rbox{1pt}{$_{\stackrel{p}{X}{}_1}$}}
%%\lieni{p}{1}
$d^{V}$\ff{s}$_{2}$} \nn \\
                                  & =\, &
\mbox{$ (-1)^{p+1}\, d^{V}($\xxi{p}{1}\inn\ $d^{V}$\ff{s}$_{2})$} \\
                                  & =:  &
\mbox{$- d^{V} \pbr{\ff{r}_{1}}{\ff{s}_{2}},$}
\end{eqnarray}
where $r=n-p$ and $s=n-q$. From (3.5) it follows that the
SN bracket of two Hamiltonian fields is a Hamiltonian
field (as in mechanics). In (3.6) one defines the
bracket operation on Hamiltonian forms which
is induced by the vertical SN
bracket of multivector fields associated with them.

{}From the definition in (3.6) it follows
\beq
\mbox{$\pbr{\ff{r}_1}{\ff{s}_2} = (-1)^{(n-r)}X_{1} \inn\ d^{V} \ff{s}_2
= (-1)^{(n-r)}X_{1} \inn\ X_{2} \inn\  \Omega $    }
\eeq
and
\beqa
deg \pbr{\ff{r}_1}{\ff{s}_2} &=& r+s-n+1, \\
%(ii) graded antisymmetry
\mbox{$\pbr{\ff{p}_1}{\ff{q}_2}$} &=& -(-1)^{g_1 g_2}
\pbr{\ff{q}_2}{\ff{p}_1},    \\
%(iv) graded Jacobi identity
\mbox{$(-1)^{g_1 g_3} \pbr{\ff{p}}{\pbr{\ff{q}}{\ff{r}}}$}
&+& \nonumber \\
\mbox{$(-1)^{g_1 g_2}
\pbr{\ff{q}}{\pbr{\ff{r}}{\ff{p}}}$}
&+& \mbox{$(-1)^{g_2 g_3}
\pbr{\ff{r}}{\pbr{\ff{p}}{\ff{q}}} $}= 0,  \\
%(iii) graded Leibniz rule
\pbr{\ff{p}}{\ff{q} \wedge \ff{r}}&=&
\pbr{\ff{p}}{\ff{q}} \wedge
\ff{r} + (-1)^{q(n-p-1)} \ff{q} \wedge
\pbr{\ff{p}}{\ff{r}} \nn \\
&+& \mbox{\it higher-order corrections},
\eeqa
where $g_1 = n-p-1$, $g_2 = n-q-1$ and $g_3 = n-r-1$.

The algebraic construction which satisfies the axioms
(3.9), (3.10)
and (3.11) without higher-order corrections,
together with the familiar properties of the
exterior product, is known as the Gerstenhaber
algebra.\cite{Gerst} Higher-order corrections in Eq. (3.11)
are composed of the terms like
\[ \frac{1}{(n-p-1)!}X^v{}\uind{i}{n-p-1}\, (\der_v{}\lind{i}{s}
\inn \dv \ff{q}) \we \der_{i_{s+1}}{}_{i_{s+2}}{}_{...}{}_{i_{n-p-1}}
\inn \ff{r}, \]
with $s=1,...,n-p-1$,
and are similar to the last term in the "Leibniz rule" for,
say, the second derivative:
$ (fg)''=f''g + fg'' + 2 f'g'$.
They  appear due to the fact that the multivector field
$\xx{n-p}$ associated with the form $\ff{p}$ does not
act on exterior forms as a graded derivation, but rather
as a graded differential operator of order $-(n-p)$
which is composed of the subsequent actions of graded derivations
of  order $-1$. The latter are given by the vector fields
$\der_v$ and $\der_{i_s}$, $s=1,...,n-p-1$, which constitute
the vertical multivector $\xx{n-p}$. The algebraic
structure given by Eqs. (3.9)--(3.11) may be called the
higher-order Gerstenhaber algebra, but in the following we will
continue to refer to it as  a Gerstenhaber  algebra, for short.

\medskip

{\sl Remark:} Strictly speaking, the space of Hamiltonian forms
is not closed with respect to the exterior product, so that
the full justification of the Leibniz rule (3.11) requires a
generalization of the above construction which admits
arbitrary horizontal forms as the dynamical
variables (see Ref. 12).
%%%%%%%%%%%%%%%%%%%%%%%%%%%%%%%%%%%%%%%%%%%%%%%%%%%%%%%%%%%%
%%%%%%%%%%%%%%%%%%ADDED LINE %%%%%%%%%%%%%%%%%%%%%%%%%%%%%%%%
%%%%%%%%%%%%%%%%%%%%%%%%%%%%%%%%%%%%%%%%%%%%%%%%%%%%%%%%%%%%%
Higher-order corrections in
(3.11) were overlooked in previous communications
(cf. Refs. 1,12).
%%%%%%%%%%%%%%%%%%%%%%%%%%%%%%%%%%%%%%%%%%%%%%%%%%%%%%%%%%%%%

\section{\hspace{-4mm}.\hspace{2mm}EQUATIONS
OF MOTION IN THE BRACKET FORM}

\hspace*{0.8cm}By analogy with mechanics, one can expect that
the equations of motion are given by the bracket with the DW
Hamiltonian function. Indeed, for the bracket of $H$ with the
$(n-1)$--form $F:=F^i\om_i$ one obtains
\beqa
\pbr{H}{F}&=&X_F\inn\ \dv H=X_F{}^a\der_a H+
X_F{}^i_a \der^a_iH \nn \\
&=&\der_i p_a^j \der^a_j F^i + \der_i y^a\der_aF^i,  \nn
\eeqa
where we have used (2.3) and (1.1). Introducing the
{\em total} (i.e. evaluated on extremals)
exterior differential
$\bd$ of a horizontal  form of degree $p$, $\ff{p}$:
\[\bd \ff{p} := \der_iz^M dx^i\we\der_M\ff{p}
=\der_iz^vdx^i\we\der_v\ff{p}+dx^i\we \der_i\ff{p}
= \bd^V F + d^{hor} F, \]
one can write the equation of motion of Hamiltonian
$(n-1)$--form $F$ as (by definition, $*^{-1}\om := 1$)
\beq
*^{-1} \bd F=\pbr{H}{F}+\der_iF^i.
\eeq

The bracket of a $p$--form with $H$ vanishes for $p<n-1$.
The equations of motion of arbitrary forms may be written in terms
of the bracket with the $n$--form $H\om$. This implies a certain extension
of the construction in Section 2. Namely, we map $H\om$ to a vector-valued
form $\xt:=\xt^v{}_i dx^i \otimes \der_v$ by
\beq
\xt \inn\ \Omega = \dv H\om,
\eeq
where
\mbox{$\tilde{X}\inn\ \Omega := X^{v}_{\cdot\, k} dx^{k}
\wedge\, (\der_{v}\inn\ \Omega) $} is the  Fr\"ohlicher-Nijenhuis inner
product. From (4.2)
it follows
\beq
\mbox{$\tilde{X}$$^a_{\cdot k}$$ = \der^a_k H,$ \hspace*{1em}
$\tilde{X}$$^i_{a k}$$\delta^k_i = -\der_a H . $}
\eeq
Substitution of the natural  parametrization
of $\xt$:
\[ \tilde{X}^v_{\cdot k} = \frac{\der z^v}{\der x^k} , \]
into (4.3) leads to the DW Hamiltonian equations (1.1).

Now, we define the bracket with $H\om$ (cf. (3.7))
\beq
\pbr{H\om}{\ff{p}}%\tilde{}
:=\xt_{H\om}\inn \dv \ff{p}
\eeq
and find that
\beq
\mbox{{\bm $d$}}\ff{p}=
\pbr{H\om}{\ff{p}}%\tilde{}
+d^{hor}\ff{p} .
\eeq
Thus, we have shown that the bracket with the DW Hamiltonian
$n$--form $H\om$ is related to the exterior differential of a
form.

\medskip

{\sl Remark:} The bracket which is naively defined in (4.4)
does not satisfy in general the axioms of a Gerstenhaber
algebra. The appropriate extension of a Gerstenhaber algebra
structure to $n$--forms is a part of the generalization
of the present construction to the forms which are not
Hamiltonian according to the definition in Section 2
(see Ref. 12). %%\cite{kanat}).

\section{\hspace{-4mm}.\hspace{2mm}TOWARDS A QUANTIZATION}

\hspace*{0.8cm}An appropriate quantization of a Gerstenhaber algebra
of exterior forms, which generalizes to field theory the
Poisson algebra of dynamical variables, may in principle
lead to certain quantization procedure in field theory.
The purpose of this section is to discuss briefly
a possible heuristic approach
to such a quantization.

We start from the observation that
\beq
\pbr{p_a}{y^b}=\delta_a^b ,
\eeq
where $p_a:=p_a^i\om_i$ is the $(n-1)$-form
which may be considered as the momentum variable
canonically conjugate to fields $y^a$.
Applying Dirac's quantization prescription
$[ \; \; , \; \; ]_{\pm}= i\hbar \pbr{ \; \; }{ \; \; }_{\pm}$,
one obtains the canonical commutation relation for
the operators corresponding to fields and the $(n-1)$-form momenta
\beq
[\hat{p}_a,\hat{y}^b] = i\hbar \delta_a^b.
\eeq
In the ``$y$-representation'' one finds the differential operator
realization
of $\hat{p}{}_a$
\beq
\hat{p}_a = i\hbar \frac{\der}{\der y^a} .
\eeq

Based on the analogy with the quantization of classical mechanics in
Schr\"odinger's representation and the observation made in Section 4
that the exterior differential is related to the DW Hamiltonian
$n$-form, one can conjecture the following form of the
covariant ``Schr\"odinger equation''
\beq
i\hbar\, d \Psi = (H\om)^{op}\Psi
\eeq
for the ``wave function'' $\Psi=\Psi(x^i,y^a)$, which depends on the
space-time and field variables which form the analogue of a
configuration space within the present formulation.

%\medskip

In the particular example of a system of scalar fields
$y^a$  interacting through the potential $V(y)$, which is given
by the Lagrangian
\[ L= - \frac{1}{2} \der_i y^a \der^i y_a - V(y) , \]
the DW Hamiltonian function takes the form
\beq
H = -\frac{1}{2} p^i_a p^a_i + V(y) .
\eeq
In terms of the $(n-1)$-form momenta variables $p_a$ the
$n-$form $H\om$ may be written as
\beq
H\om = \half *p_a \we p^a + V(y)\om ,
\eeq
where $*p_a=-p_a^idx_i$
(the Minkowski metric in the $x$-space is assumed). The realization of
the operator corresponding to the non-Hamiltonian one-form $*p_a$
\beq
\widehat{*p}_a  = *\, \hat{p}_a
\eeq
is suggested by the quantization of the bracket
\beq
\pbr{*p_a}{y^b\om_i}= - \delta_a^b dx_i
= * \pbr{p_a}{y^b\om_i},
\eeq
which may be calculated either with the help of the Leibniz rule (3.10)
or within a more general scheme,\cite{kanat}
where one
associates arbitrary horizontal forms with the differential operators
on exterior algebra which are represented by the multivector-valued forms
instead of the multivectors as in the case of Hamiltonian forms.

Furthermore, the classical identity $\om=*1$ suggests that $\hat{\om}= *$
and, therefore, one can write
\beq
(H\om)^{op}=* (-\frac{\hbar^2}{2} \lapl + V(y)) =:*H^{op} ,
\eeq
where $\lapl := \der^a \der_a$ is the Laplace operator in a field
space. Thus, the Schr\"odinger equation (5.4) may also be written as
\beq
i \hbar *^{-1} d\Psi = H^{op} \Psi.
\eeq
Evidently, this equation makes sense only if the wave function $\Psi$
is a nonhomogeneous horizontal form. In the simple case of the
DW Hamiltonian operator (5.9), which does not depend explicitly on
the space-time coordinates, one can take into account only the
zero- and $(n-1)$-form contributions,  so that
\beq
\Psi=\psi_0(x,y) + \psi^i(x,y)\, \om_i.
\eeq
Substituting (5.11) into (5.10), one obtains the component form
of our
Schr\"odinger equation:
\beqa
 i\hbar \der_i \psi^i &=& H^{op} \psi_0   , \\
 -i\hbar \der_i \psi_0 &=& H^{op} \psi_i .
\eeqa
The integrability condition of this set of equations is
\beq
\delta \Psi = 0.
\eeq
By a straightforward calculation, one can derive from (5.12)
and (5.13) the following conservation law
\beq
\der_i [\psib_0 \psi^i + \psi_0 \psib{}^i  ]
= - \frac{i\hbar}{2} \der_a[\psib_0 \derts_a \psi_0 -
\psib^i \derts_a \psi_i ]          .
\eeq
If one assumes a sufficiently rapid decay of the wave function
$\Psi (x,y)$ for large values of fields $|y| \to \infty $,
by Gauss' theorem one obtains
\beq
\der_i \int dy\, [\psib_0 \psi^i + \psi_0 \psib{}^i  ] = 0  .
\eeq
Thus, the current
\[ j^i := \int dy\, [\psib_0 \psi^i + \psi_0 \psib{}^i  ] \]
is the conserved space-time current of the theory.
It suggests the inner product of nonhomogeneous forms $\Psi$,
which one needs for the calculation of quantum theoretical
expectation values.

%\medskip

The covariant Schr\"odinger equation may be solved
by means of the separation
of field and space-time variables.
Namely, let us write
\beq
\Psi(x,y)=\Phi(x)f(y),
\eeq
where $\Phi(x)$ is a nonhomogeneous form with the components
depending on $x$:
\beq
\Phi(x):=\phi_0(x)+\phi^i(x)\omega_i
\eeq
and $f(y)$ is a function of field variables. Substituting this Ansatz
into the Schr\"odinger equation (5.10), we arrive at
the eigenvalue problem for the DW Hamiltonian operator
\beq
H^{op} f = \kappa f ,
\eeq
and the equation on $\Phi(x)$:
\beq
 i\hbar *^{-1}d \Phi(x) = \kappa \Phi(x).
\eeq
{}From the latter equation it follows that
\beq
\Box \phi_0 =\frac{\kappa^2}{\hbar^2} \phi_0, \quad
\phi_i = -\frac{i\hbar}{\kappa} \der_i \phi_0.
\eeq
The solutions of (5.19) and (5.21) provide us with a basis for
decomposition of an arbitrary solution of the
covariant Schr\"odinger equation.

\medskip
         %%%%%\newpage

{\sl Remarks:}

1. The canonical bracket (5.1) belongs to the subalgebra of
zero- and $(n-1)$-forms of a Gerstenhaber algebra of
dynamical variables. The other
canonical brackets from this subalgebra are
\beqa
\pbr{p_a^i}{y^b\om_j}&=& \delta^b_a \delta^i_j , \\
\pbr{p_a}{y^b\om_i}&=& \delta_a^b \om_i .
\eeqa
Quantization of these three brackets is a part of the problem
of quantization of the center of a Gerstenhaber algebra, which
is
formed by the forms of the kind $p^i_a dx^{...}$ and $y^a dx^{...}$,
where $dx^{...}$ denotes the basis elements of Grassmann algebra of
horizontal forms. The question as to which subalgebra of a
Gerstenhaber algebra of dynamical variables should or can be
quantized remains open and deserves the same careful study as
the similar question concerning the quantizable subalgebra of
the Poisson algebra of observables in mechanics. The minimal
subalgebra is that of $(n-1)$-forms and the canonical bracket from
this subalgebra is given by (5.23).
Its quantization rather than a
quantization of (5.1) gives rise to the operator realization of $p_a$
in (5.3). The quantization of the subalgebra of zero- and $(n-1)$-forms
with the canonical brackets (5.1), (5.22) and (5.23)
leads to the problem of
realization of the operator $\hat{p}{}_a^i$ which would be
consistent with the realization of $\hat{p}_a$ and the
requirement $\hat{p}_a = \hat{p}{}_a^i \co \hat{\om}_i$, as
well as to the problem of the proper realization of the operation $\co$ of
the
multiplication of quantum operators. When quantizing the
(centre of the) Gerstenhaber algebra, the latter problem is that
of the proper realization of the quantized wedge product, which is
in this case a generalization of the Jordan symmetric product of
operators in quantum mechanics.

2. The quantization of the bracket (5.1) leads to the operator
realization of the $(n-1)$--form $p_a$ which is the $0$-form.
In general, the form degree of the operator corresponding to
a dynamical variable is different from the classical form degree
of the latter. This gives rise to an additional problem of which
degree should define the graded products of operators
which correspond to the exterior product and the quantized
Poisson bracket respectively.

3. It is interesting to note that the realization (5.3) is not
consistent with the classical property
$dx^i \we p_a = (-)^{n-1} p_a \we dx^i$
which one may require to be also fulfilled on the quantum level.
This may be achieved if the quantization prescription is modified
in such a way that
\beq
[ \; \; , \; \; ]_{\pm}= \gamma \hbar \pbr{ \; \; }{ \; \; }_{\pm}
\eeq
where $\gamma$ denotes the imaginary unit corresponding to the
Clifford algebra of the $n$-dimensional space-time
%\cite{Hestenes, Rashevski, ...}
 over which a field theory
under quantization is formulated. In Minkowski space-time
$\gamma := \gamma_0 \gamma_1 \gamma_2 \gamma_3$. In the case
of mechanics ($n=1$) $\gamma=i$ and the above quantization
prescription reduces to that of Dirac.
The quantization according to (5.24)
leads to the realization
\beq
\hat{p}_a = \gamma \hbar \der_a
\quad {\rm and} \quad
\widehat{dx}{}^i \we=\gamma^i \we ,
\eeq
where $\we$ on the right hand side
denotes the graded symmetrized Clifford product. Correspondingly,
the wave function may be considered as taking values in the
Clifford--K\"ahler algebra of nonhomogeneous forms which
corresponds
to the $n$-dimensional space-time (see e.g. Ref. 13 %%\cite{dk}
and the
references  quoted there). The latter
reduces to complex numbers in the case of mechanics.
This quantization prescription leads to the same realization of the
DW Hamiltonian operator as in (5.9). However, in general, it is not clear
which quantization prescription is more appropriate both physically
and mathematically for the quantization of the suitable
``quantizable'' subalgebra of a Gerstenhaber algebra of
forms--dynamical--variables in field theory.

4. The elements of quantum theory presented above
possess the basic features of a quantum description of
dynamics  and its
connections with the structures of classical mechanics.
These elements  are easily seen to reduce to the corresponding
elements of quantum mechanics at $n=1$.
In this sense at least our formulation may be viewed as
an approach to the quantum description of fields.
Establishing  the possible links with the known approaches
and results in quantum
field theory and a physical interpretation of the present
formulation  poses many conceptual questions and needs a further
study which we hope to communicate elsewhere.
In particular, it would be interesting
to understand a possible
relation of our nonhomogeneous form-valued wave function
$\Psi (x,y)$ to
the Schr\"odinger wave functional $\Psi (t,[y({\bf x})])$
and of our covariant Schr\"odinger equation to the functional
Schr\"odinger equation.

%Conclusions

\medskip

{\bf Acknowledgements.} I thank Prof. A. Odzijewicz and the organizers
for inviting me to present this talk. I acknowledge useful
discussions with F. Cantrijn, M. Gotay, M. Modugno and
J. \'Sniatycki during the time of the workshop. Thanks to
Z. Oziewicz for several inspirating discussions on the subject
of this paper and encouragement.


\begin{thebibliography}{99}
\newcommand{\bib}[1]{\bibitem{#1}}
\footnotesize

\bib{Kan} I.\,V. Kanatchikov, On the canonical structure of
the De Donder-Weyl covariant Hamiltonian formulation of field
theory I. Graded Poisson brackets and equations of motion,
PITHA 93/41 (November 1993) and hep-th/9312162

\bib{Rund} H. Rund, ``The Hamilton-Jacobi Theory in the Calculus of
Variations'',  van Nostrand, Toronto  (1966)

\bib{Kastrup83} H. Kastrup,
 Canonical theories of Lagrangian dynamical systems in physics,
{\em Phys. Rep.} 101:1 (1983)

\bib{Binz} E. Binz, J. \'{S}niatycki, H. Fisher, ``Geometry of
Classical Fields'', North-Holland, Amsterdam (1989)

\bib{Kij ea} J. Kijowski,
 A finite dimensional canonical formalism in
the classical field theory,
{\em Comm. Math. Phys.} 30:99 (1973); \\
J. Kijowski,  Multiphase spaces and gauge in the
calculus of variations,
{\em Bull. de l'Acad. Polon. des Sci.,
S\'er sci. math., astr. et Phys.} XXII:1219 (1974); \\
J. Kijowski and W. Szczyrba,
 A canonical structure for classical field theories,
{\em Comm. Math. Phys.} 46:183 (1976)


\bib{Gotay} M.J. Gotay,
 An exterior differential systems approach to the Cartan form,
{\em in}: ``G\'{e}om\'{e}trie Symplectique $\&$
Physique Math\'{e}matique'', P. Donato, C. Duval et al. (eds.),
Birkh\"{a}user,
Boston (1991) \\ %p.160
M.\,J. Gotay, A multisymplectic framework for classical field theory
and the calculus of variations I. Covariant Hamiltonain formalism,
{\em in}: ``Mechanics, Analysis and Geometry:
200 Years after Lagrange'',  M. Francaviglia (ed.), North Holland,
Amsterdam  (1991) \\ %p. 203
M.\,J. Gotay,   A multisymplectic framework for classical field theory
and the calculus of variations II. Space + time decomposition,
{\em Diff. Geom. and its Appl.} 1:375 (1991)

\bib{Gimmsy} M.\,J. Gotay, J. Isenberg, J. E. Marsden, R.
Montgomery, J. \'{S}niatycki and Ph. B. Yasskin:
 Momentum maps and
classical relativistic fields: The Lagrangian and Hamiltonian
structure of classical field theories with constraints,
preprint, Berkeley (1992)

\bib{Crampin} J.\,F. Cari\~nena, M. Crampin, L.\,A. Ibort,
On the multisymplectic formalism for first
order field theories,
{\em Diff. Geom. and its Appl.} 1:345 (1991)

\bib{Abr+Marsden} R. Abraham and J.\,E. Marsden,
``Foundations of Mechanics'',
2nd ed., Benjamin and Cummings, N.Y. (1978)

\bib{tulcz} W.\,M. Tulczyjew, The graded Lie algebra of
multivector fields and the generalized Lie derivative of forms,
{\em Bull. de l'Acad. Polon. sci., S\'er sci. math., astr. et phys.}
XXII:937 (1974)

\bib{Gerst} M. Gerstenhaber,
 The cohomology  structure of an associative ring,
Ann. Math. 78:267 (1963); \\
M. Gerstenhaber and S.\,D. Schack,
Algebraic cohomology and deformation theory,
{\em in}: ``Deformation Theory of Algebras
and Structures and Applications'', M. Hazewinkel and M.
Gerstenhaber (eds.), Kluwer, Dordrecht (1988);\\
B.\,H. Lian and G.\,J. Zuckerman,
New perspectives of the BRST-algebraic structure of
string theory,
{\em Commun. Math. Phys.} 154:613 (1993)

\bib{kanat} I.\,V. Kanatchikov,
Basic structures of the covariant canonical formalism for fields
based on the De Donder-Weyl theory,
preprint PITHA 94/17 and hep-th/9410238; \\
I.\,V. Kanatchikov,
On the finite dimensional covariant Hamiltonian formalism
in field theory, {\em in}:
``New Frontiers in Gravitation'', R. Santilli and G. Sardanashvily (eds.),
Hadronic Press, Palm Harbor (1995) (to appear)

\bib{dk} P. Becher and H. Joos,  The Dirac-K\"ahler equation
and fermions on the lattice, {\em Z. Phys. C} 15:343 (1982),\\
I.\,M. Benn and R.\,W. Tucker,  Fermions without spinors,
{\em Comm. Math. Phys.} 89:341 (1983),\\
N.\,A. Salingaros, G.\,P. Wene,  The Clifford algebra of differential
forms, {\em Acta Appl. Math.} 4:271 (1985).
\end{thebibliography}
\end{document}